\documentstyle[aps,epsfig]{revtex}
\begin{document}

\draft
\title{Observation of the Mott Effect in Heavy Ion 
Collisions}

\author
{P.\
Bo\.{z}ek$^{1,2}$\footnote[1]{E-mail~:~bozek@solaris.if
j.edu.pl
}, P.\ Danielewicz$^{3}$,
K.\ Gudima$^{4,5}$,     and   M.\ P\l{}oszajczak$^{4}$  
  }

\address{$^{1}$  Yukawa Institute for Theoretical 
Physics, Kyoto
University, Kyoto 606-01 , Japan \\
$^{2}$ Institute of Nuclear Physics, 31-342 Krak\'{o}w, 
Poland  \\
$^{3}$ National Superconducting Cyclotron Laboratory 
and Department
of Physics and Astronomy, Michigan State University, 
East Lansing,
Michigan 48824, USA  \\
$^{4}$~ Grand Acc\'{e}l\'{e}rateur National d'Ions 
Lourds (GANIL),
BP 5027,  F-14021 Caen Cedex, France \\ and \\
$^{5}$~ Institute of Applied Physics, Moldova Academy 
of Sciences,
277028 Kishineu, Moldova    }

\date{\today}

\maketitle

\begin{abstract}
\parbox{14cm}{\rm

Possibility of observing the Mott momentum in the
distribution of the deuterons produced in the process
$p + n  \rightarrow d + \gamma$, in the first stage of 
a
nuclear reaction is presented. The correlation of a 
hard
photon  with a deuteron  allows to
select those deuterons produced at the beginning of a 
reaction.
\smallskip\\}
\end{abstract}

\bigskip
Corresponding author :\\
M. Ploszajczak\\
GANIL, BP 5027\\
F-14076 Caen Cedex 05\\
France

\vfill
\newpage


Recent analysis of the hard photon production in heavy 
ion collisions
shows the importance  of the two-body
process $\pi + N \rightarrow N +  \gamma$
in  the description of  the high energy part
of the photon spectrum observed experimentally by the 
TAPS
collaboration \cite{gp}. The inclusion of the processes
$p + n \rightarrow d + \pi $ and $p + n \rightarrow d + 
\gamma$
seems to be necessary in order to improve, 
simultaneously, the
description of the data for the pion and photon 
production. The two body
phase-space and a weak dependence of the cross section
$\sigma_{n  p \rightarrow d  \gamma}$ on energy make 
the $p+N
\rightarrow d + \gamma$ process
important for the high energy part of the photon 
spectrum.
The deuteron in the final state is not Pauli-blocked.
If we allowed
for all  deuteron momenta, that process would be the 
dominant
source of
high energy photons ($E_\gamma > 30$MeV).  However, the 
process
$p + n  \rightarrow d + \gamma$~ requires the
existence of the bound state of the final deuteron. 
That means,
that
the deuteron momentum  $p_d$~ must be above the Mott 
momentum
$p_{Mott}$~ \cite{ogmott,ogmott1,bd},
if the  influence of the surrounding nucleons is taken 
into account.
In particular, deuterons with low momentum cannot exist 
at a normal
nuclear density. Typically, one finds the condition 
$p_d/2 >
p_{Mott}/2
\simeq 1.2 p_F$ for the existence of a deuteron around 
the
normal nuclear density~\cite{bd}.

The process involving a~deuteron was not previously 
discussed
in the context of hard photon production in heavy ion
reactions~\cite{review}. Thus
it seems important to test the relevance of that 
process in a
more specific way, using
the $d - \gamma$ correlation. The correlation to a hard 
photon allows us
to select the deuterons from this 2-body process.
Additional
condition on the minimal energy of the photon or the 
maximal energy
of deuteron allows to select the deuteron production
at the early stage of a nuclear reaction.
Another way of producing deuterons involves 3-body 
collisions\cite{bd}
and it also utilizes as a  basic ingredient
the value of  the Mott momentum.  Below,
we show how in  a more direct way
the Mott effect  can be tested in heavy ion
reactions using the simpler $p + n \rightarrow d + 
\gamma$~
process, where both the photon and the deuteron in the 
final
state can be detected experimentally.

To understand basic elements of the production process,
the $d - \gamma$~ correlations will be studied in a
simple first-chance collision model (FCCM). In this 
model of
the initial phase of a heavy ion reaction,
the initial momentum distribution of nucleons, that 
limits
the deuteron momenta, is given by the two Fermi
spheres\footnote{This corresponds to the $T = 0$~ 
assumption
for the nuclear medium in which the
$p + n \rightarrow d + \gamma$~ process takes place.
In consequence, the validity of FCCM is restricted to 
the
early stage
of a reaction.}.  For that distribution, the deuterons 
are
emitted predominantly
at $90^\circ$ in the nucleus-nucleus center of mass 
(c.m.).
In describing the $p + n \rightarrow d + \gamma$~ 
process, one
needs to take into the account
the interaction of the nucleon pair (the deuteron) with 
the
surrounding
medium. As shown in  \cite{ogmott,ogmott1}, the 
formation of bound
two-particle states is strongly suppressed at large 
density
and/or low temperature values due to Pauli blocking 
which
limits, additionally, the deuteron momenta. To 
implement
this effect, we parametrize the allowed region for the 
bound
deuterons as follows~\cite{bd}:
\begin{equation}
\label{motteq}
\int f(\frac{1}{2}{\bf p_d}- {\bf p}) \, |<{\bf p}| 
\phi >|^2
\, \frac{d^3p}{(2\pi)^3} \ <
0.2~~~~ \ ,
\end{equation}
where $<{\bf p}|\phi>$ is the deuteron wave function
and $f(p) $ is the  momentum distribution of the 
surrounding nucleons.
In~crude terms, the deuterons can be formed if:
$p_d/2$~\raisebox{-.5ex}{$\stackrel{>}{\scriptstyle\sim
}$}~$
p_F$, and the direction of $p_d$ is transverse to the
collision axis in the nucleus-nucleus center of mass.  
This is
illustrated in Fig.~1;
the deuteron half momentum should find itself 
in-between the
Fermi
spheres of the initial nuclei.  The contributing 
nucleons must
have large transverse momenta, in the vicinity of the 
Fermi
momentum.
The minimal transverse momentum of the emerging 
deuteron as a
function of the energy of the collision is shown in 
Fig. 2.
We see that around $E_{lab} = 140$MeV/A the lower limit 
on
deuteron momentum disappears.  Also at  around 
$50$MeV/A, the
kinematical
restriction does not allow for the production of a  
$d-\gamma$
pair ($E_\gamma > 30$MeV). In practice, the interesting 
energy range
for the study of the dependence of the Mott momentum on 
the incident
energy is $70-140$MeV/A. It should be noted that the 
value of the
minimal Mott momentum depends relatively strongly on 
the chosen
value for the Fermi momentum. In particular, if the 
reaction takes
place at the nuclear surface, the Fermi momentum, 
depending on
the
local density, is low.  The curves in Fig.~2, which 
correspond
to
different values of $p_F$~ representing the effective 
Fermi momentum
for either a lighter or a heavier  nucleus or for a low 
density
in the nuclear surface, differ substantially.
The results presented below were obtained for 
$p_F=210$MeV/c.

The probability of the production of the $d - \gamma$~ 
pair
per participant can be described in the first chance
collision picture by~\cite{bertsch77}:
\begin{eqnarray}
\frac{dP_{\gamma d}}{d^3 p_\gamma \ d^3p_d} & = &
\left( \frac{1}{2}
\int \int \frac{d^3p_1}{(2\pi)^3} 
\frac{d^3p_2}{(2\pi)^3}
\, f({\bf p}_1) \, f({\bf p}_2) \,
\frac{|{\bf p}_1 - {\bf p}_2 |}{m} \, \frac{d 
\sigma_{np
\rightarrow \gamma d}}{d{\bf p}_\gamma d{\bf p}_d} \,
\Theta({\bf p}_d,{\bf p}_{mott}) \right)  \times
\nonumber \\  \nonumber  \\
& {\times} & \left(
{<\sigma_{nn}>_{eff} \, \frac{p_{coll}}{m} \,  
\rho_0^2}
\right)^{-1}  \ ,
\end{eqnarray}
where  $\Theta$ is 1 or 0 depending whether a bound 
deuteron
state of momentum
${\bf p}_d$ exists or not, and where $\rho_0$ and
$\sigma_{nn}=40mb$ are the nuclear density
and the average nucleon-nucleon cross section in 
medium,
respectively.  The first
factor on the right hand side of Eq.~(2) is the rate
of
production of $d - \gamma$ pairs per unit volume and 
unit time.
Dividing that rate by the collision rate,
we get an
estimate for the number of $d - \gamma $ pairs from the 
first
stage of
a reaction per participant; $<\sigma_{NN}>_{eff}$ is
the effective Pauli blocked cross section 
\cite{bertsch77} and
${p_{coll}}/m$ is the relative velocity of the two 
nuclei.
Notice that, since we deal with the non-isotropic
momentum distribution at the beginning of a reaction, 
the Mott
momentum depends on the direction of the deuteron 
momentum, unlike in
nuclear matter.  Since the lowest Mott momentum is in 
the
transverse direction (see Fig. 1), in that direction
deuterons will be emitted predominantly.

For our calculations,
we have parameterized the experimental data
for the cross section of the deuteron breakup process, 
compiled
in Ref. 7, with
\begin{eqnarray}
\label{eq3}
\sigma_{\gamma d \rightarrow p n } [mb]
& = &
\frac{32.3}{E_\gamma^{1.39}}  \ \ , \ \ E_\gamma < 50 
MeV \nonumber \\
 & = & \frac{7.72}{E_\gamma^{1.01}} \ \ , \ \ E_\gamma 
> 50 MeV \ ,
\end{eqnarray}
where $E_\gamma$ is incident photon energy in MeV.  
The~cross
section for the deuteron production is then given by
\begin{equation}
\sigma_{p n \rightarrow \gamma d}(\sqrt{s})
= \frac {3}{2} \, \frac{s-4m_N^2}{s} \, \sigma_{\gamma 
d
\rightarrow p n} \ .
\end{equation}
The momentum distribution of the produced deuteron
and photon in the c.m.\ of the colliding $p - n$ pair 
is taken
as isotropic.

The resulting probability distribution as a function of 
the relative
angle between the photon and the deuteron,
in the nucleus-nucleus c.m.,
is shown in Fig. 3. We see that the $d - \gamma$ pair 
is not
quite emitted back to back in that frame.  The  angular
distribution in the relative
angles is broad around 180$^\circ$, with a tail 
extending down
to very small relative angles.  This is, of course, due
 to the fact that the nucleus-nucleus c.m.\  does not 
coincide
with the c.m.\ of contributing nucleons;
the relatively low number of back
to back pairs stems from imposing the condition 
$E_\gamma
>30$MeV~ in the
nucleus-nucleus c.m.\footnote{The Mott condition 
(\ref{motteq}) implies
that the deuterons are mostly emitted in the
direction of c.m.\ momentum of the colliding pair, so 
that the
momentum
of the deuteron is increased when going from the
nucleus-nucleus c.m.\ to the nucleon-nucleon c.m.
The contrary is true for most of the photons.}

The momentum distribution of the deuterons emitted
in the reaction $p + n \rightarrow d + \gamma$ is shown 
with the solid
line in the upper part of Fig. 4.  Consistent with our
expectations we find that the deuteron distribution is 
bound from
below by the value of the Mott momentum and from above 
by the
kinematical restrictions. This shape of the deuteron 
momentum
distribution is very different from the usual spectrum 
of light
particles emitted in nuclear reactions. The deuterons 
emitted  in
the later stage of a reaction, coming mostly
from the 3-body process\footnote{In the latter stage of 
a
reaction, the momentum distribution is almost
isotropic. The production of a deuteron from the 2-body 
process
requires in that case the collision of two nucleons 
from the tail of
the momentum distribution
so that total momentum of the pair is larger than
$p_{Mott} \simeq 2.4 p_f$.}, have
momenta that can extend to lower values. Also the 
angular
distribution
would be very different for all deuterons than for 
deuterons
emitted in the first stage of a reaction.
The deuterons produced in the
2-body process have momenta directed transversely in 
the
nucleus-nucleus c.m.\ (Fig. 5).  This reflects the 
existence of
the gap
between the two Fermi spheres which helps the formation 
of a bound
deuteron (see Fig. 1). The 2-body process predicts a 
very narrow range of
angles for the direction of the  momentum of the 
deuterons observed
in the correlation with a hard photon.  In the
laboratory frame of reference that range is  around $ 
40^\circ$
in the forward direction for a symmetric system,
i.e.\ for nuclei having the same effective Fermi 
momentum.

From the observed distribution of deuterons produced in
the correlation with a hard photon, we can deduce the 
Mott
momentum for the initial phase
space configuration in a collision. The very narrow 
angular
distribution of deuterons produced in the first stage 
of the
collision  can be used to reduce the background from 
deuterons
produced in the 3-body collisions at latter stages of 
the
reaction.

The anticipated background for the observation of the 
photon
deuteron
correlation involves mainly the bremsstrahlung photons 
and the
deuterons produced in the 3-body processes. These can 
form
background pairs either
with a correlated $\gamma$ or deuteron, or with 
themselves.
  The bremsstrahlung of photons is known to take place 
at very
low rates for high
energy photons \cite{gp,gan}, comparable to the rates 
of
production for the $d - \gamma$~ channel. This 
background can
be
further reduced if a supplementary condition on the 
photon energy is
taken. The spectrum of photons in the nucleon-nucleon
c.m.\ frame is harder than the spectrum in the 
nucleus-nucleus
c.m.\ frame.
Thus we propose to impose energy cuts on the energy of 
the photon in
the $\gamma - d$ i.e.\ nucleon-nucleon c.m.\ and not on
$E_\gamma$ in the overall c.m.,
except for the 'minimal' condition: $E_\gamma > 30$MeV.
Such a procedure does
not reduce dramatically the cross section for the 
production of the
the $d - \gamma$~ pair (see the dashed curve in Fig. 
4).
On the other hand, the bremsstrahlung cross
section decreases by several orders of magnitude when 
changing the
energy of the photon in the $\gamma - d$ frame from 30 
MeV to
$60$MeV. Thus the background
coming from the bremsstrahlung photons can be neglected 
in comparison
to the background from the deuterons produced in 3-body 
collisions.
The deuterons produced in 3-body collisions in the 
first stage of the
reaction have the same angular distribution as the 
deuterons produced
by the 2-body mechanism. Accordingly,
they could, with a photon coming from
the $d - \gamma$ channel\footnote{Since, through 
suitable
conditions on energy, the yield of bremsstrahlung 
photons can
be made insignificant, we
do not take them into account.} be mistaken as a 
correlated
pair produced in the same nucleon-nucleon collision.

The probability of the production of a deuteron from 
the 3-body
process, per 2-body collision  participant,
in the impulse approximation can be written 
as~\cite{3n-dn,bd}~:
\begin{eqnarray}
\frac{dP}{d^3 p_d} & =& \frac{3}{4}
\int \frac{d^3 p_1}{(2 \pi)^3}\frac{d^3 p_2}{(2 \pi)^3}
\frac{d^3 p_3}{(2 \pi)^3} \, d\Omega \,
 f({\bf p}_1) \, f({\bf p}_2) \, f({\bf p}_3) \, 
(1-f({\bf
p}_1^{'})) \, \frac{|{\bf p}_1 - {\bf p}_2|}{m} 
\nonumber
\\ & & \times \frac{d\sigma_{NN}}{d\Omega} \,
 |<{\bf p}_2^{'}-{\bf p}_3|\phi>|^2 \,
\Theta({\bf p}_2^{'}+{\bf p}_3,{\bf p}_{Mott}) \left(
<\sigma_{NN}>_{eff} \, \frac{p_{coll}}{m} \, \rho_0^2
\right)^{-1}  \ .
\end{eqnarray}
In  the bottom part of Fig. 4,
we show results for the probability distribution of
deuterons produced in the 3-body process per 2-body 
collision.
We see that the yields are higher than for the 2-body 
process, but
are of the order of $10^{-7}$~.
The probability for a statistical correlation is 
proportional to the
product of the probabilities , thus very small, but 
also to the square
of the number of participants, unlike the probability 
for true
correlations which is proportional to the number of 
participants.
Also deuterons produced at later stages of the 
collision would, at some
rate, populate the same angular region as the deuterons 
produced
in the first collisions, increasing the deuteron 
background.
However, we expect that
the 2-body process : $p + n \rightarrow d + \gamma$~
, can be extracted by the simultaneous hard
photon and deuteron observation, since it leads to a
very special momentum distribution of deuterons,
which could be observed even if
some non-negligible background is present.
The deuteron distribution obtained in
this manner would tell us about the Mott mechanism in 
nuclear
matter. The photon spectrum obtained in correlation 
with a
deuteron can be  compared to
the total photon yield in order to show directly which 
part of the
spectrum is due to the photons produced in the $d - 
\gamma$~
channel. This
channel of  hard photon production is important 
especially at energies
above $140$MeV\cite{gp}~. Finally, it should be noted 
that
 the absorption of the deuterons is quite substantial,
unlike for the absorption of photons.  It can be 
estimated,
taking
$\sigma_{dN}^{ine} \simeq 2 \sigma_{NN}$, that only 
about $20\%$
of the produced deuterons survive in the reaction of 
medium
size
nuclei. Similar absorption rates are estimated in 
microscopic
calculation \cite{bd}. However, the same is true also 
for the background
deuterons, so the observation of the Mott effect in 
nuclear
matter could be possible. We estimate that, after 
accounting
for the absorption of
deuterons, the probability of the production of a 
$d-\gamma$
pair is
$10^{-6}$ per participant in the collision of medium 
size
nuclei at energy $90$MeV/A.
 It should be noted that the deuteron nucleon
cross section in medium is not well known.  One
calculation~\cite{rostdn} gave a~large breakup
cross section for a weakly bound deuteron, at momenta 
close
to the Mott momentum.
Another important effect may be the elastic scattering 
of the
produced deuterons. Thus,
we expect a broadening of the observed angular 
distributions
of deuterons and of the angular spectrum of the $d - 
\gamma$~ pairs.
Finally, important could be the deuteron absorption due 
to
the deuteron dissociation. The process may occur if the
deuteron enters a higher density region, while 
traveling in the
nuclear medium. If, in the higher density
region the deuteron is not  bound, it would dissociate. 
Similar
process can occur if the gap between the two Fermi 
spheres (Fig. 1)
becomes filled in the course of the reaction before the
deuteron
escapes the nuclear medium. However, the dissociation 
mechanism is
less important than the deuteron breakup by the 
nucleon-deuteron
collisions.

In summary, we have proposed the observation of the 
correlation of a
hard photon with a deuteron produced in heavy ions 
collision.
In the range of incident energies  70 MeV/A$ <E_{lab} < 
$
140 MeV/A, the available phase space for the bound 
deuterons
is strongly reduced by the Mott effect. This may allow 
for the
experimental observation of the Mott momentum for 
deuterons.
The experimentally estimated value of the Mott momentum 
could be
used as a phenomenological parameter in the  
calculations
of heavy ion collisions
involving the formation of a deuteron in the final 
state.
It would  be also  interesting to compare
the observed conditions  of the in medium deuteron 
formation
 to the theoretical calculations of the nuclear
medium influence on the deuteron formation.
The observation of the $d-\gamma$ correlated pairs 
would
represent a direct measurement of the photon yield 
produced in the
reaction $ p + n \rightarrow d + \gamma$ , which is 
believed to be
an important component of the total hard photon yield
in intermediate energy nuclear collisions.
The observation of the $d - \gamma$~ correlation gives, 
for the
first time, the
possibility to estimate the percentage of the hard 
photon yield from a
definite channel.
The production of deuterons in the first stage of 
a~reaction
is
determined by the allowed momenta for the deuterons 
(Fig. 1).
The experimental observation of the correlation would 
tell us
on
the mechanism of the deuteron formation in the nuclear 
medium.

The main subject of this work was the implication of 
the Mott
condition for the creation of a bound deuteron in 
nuclear medium
on the observed spectra of deuteron. However, it should 
be noticed
that the reaction discussed in this work:
$p + n \rightarrow d + \gamma$~,
seems to be important also for the description of the 
hard photon
spectra. This channel of the photon production
 has not been up to now addressed in the
microscopic calculations of heavy ion collisions.

\vspace{25mm}

\noindent
{\bf Acknowledgments}
P. Bo\.{z}ek would like to thank  YITP for the 
hospitality
extended to him. P. Bo\.{z}ek, P. Danielewicz, and K. 
Gudima
gratefully acknowledge the support of GANIL where this 
work started.
This
work was partially supported by the National Science 
Foundation
under Grant PHY-9605207.



\newpage
\begin{figure}
\begin{center}
\epsfig{file=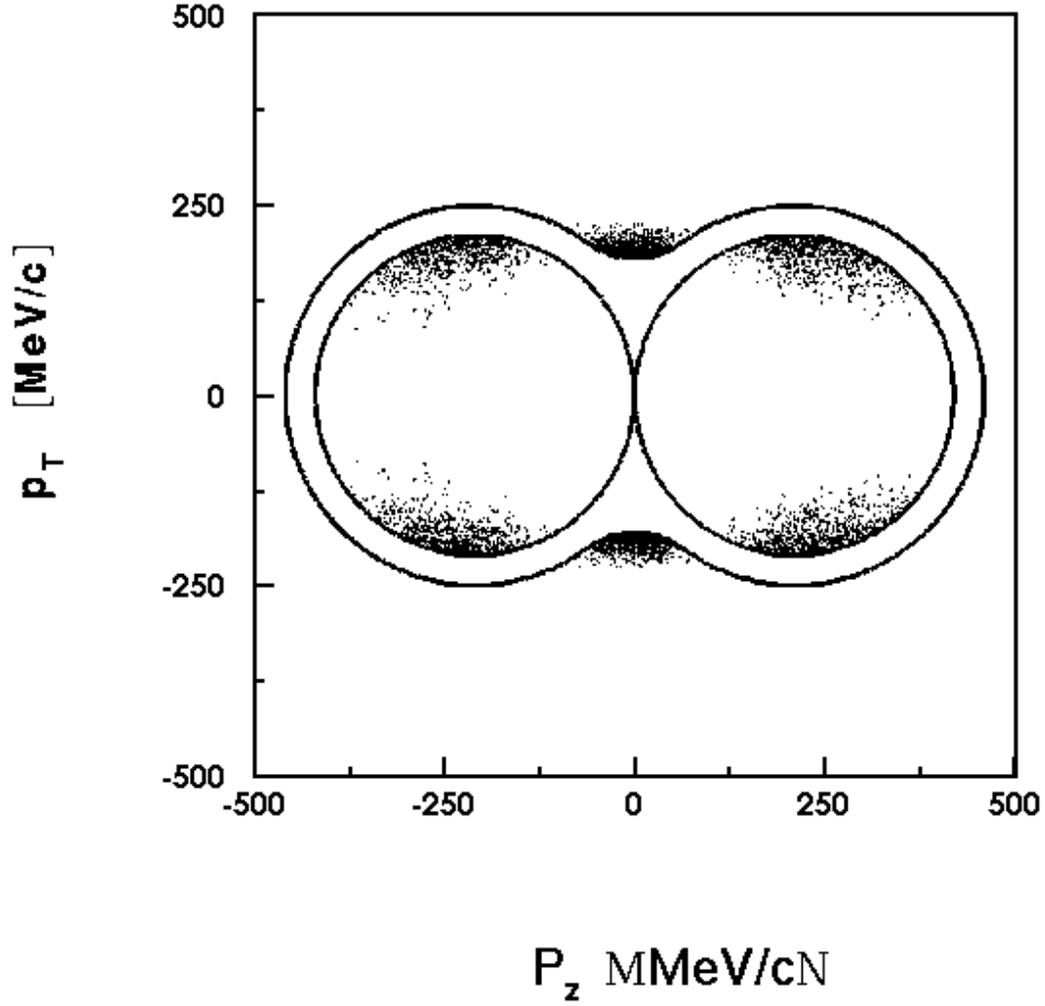,width=0.86\textwidth}
\vspace{2.5cm}
\caption{Distribution of initial longitudinal and transverse 
momenta of
colliding nucleons (the two intersecting Fermi spheres) 
for
$p_F=210$MeV at beam energy of $90$MeV/A.
The dots inside the spheres represent the momenta of 
the nucleons
that have produced a~deuteron in the 2-body process. A 
strong
reduction of the cross section, in comparison to the 
free case,
is expected due to  the restriction on the phase space 
in the
initial stage.
Half of the deuteron momentum must be located outside 
the outer
line (representing half of the angle-dependent Mott 
momentum).
The points in the outside region represent half of the 
momenta of
deuterons produced in correlation with a photon 
($E_\gamma > 30$MeV) .}
\end{center}
\end{figure}

\newpage
\begin{figure}
\begin{center}
\epsfig{file=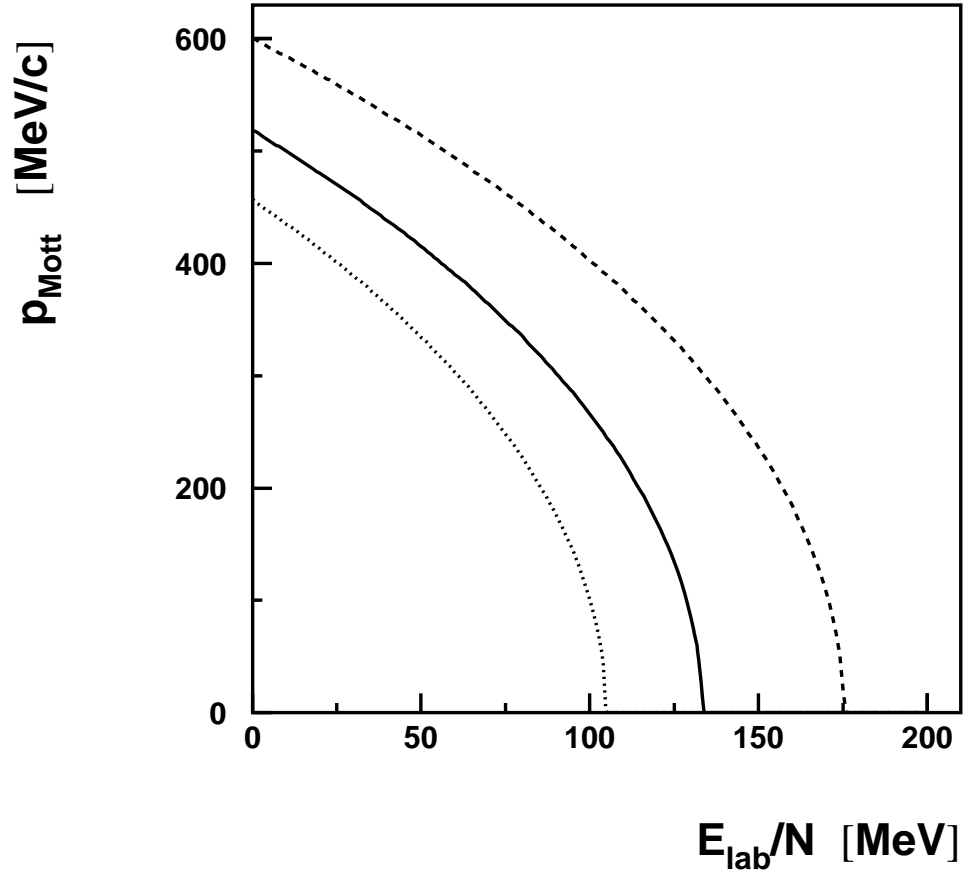,width=0.86\textwidth}
\vspace{0.5cm}
\caption{
The lowest value of the Mott momentum (corresponding to 
the
transverse direction), as a function of the collision 
energy.
The dashed, solid, and dotted lines are calculated for 
the
Fermi momentum $p_F=180$, $210$, and  $250$MeV/c, 
respectively.}
\end{center}
\end{figure}

\newpage
\begin{figure}
\begin{center}
\epsfig{file=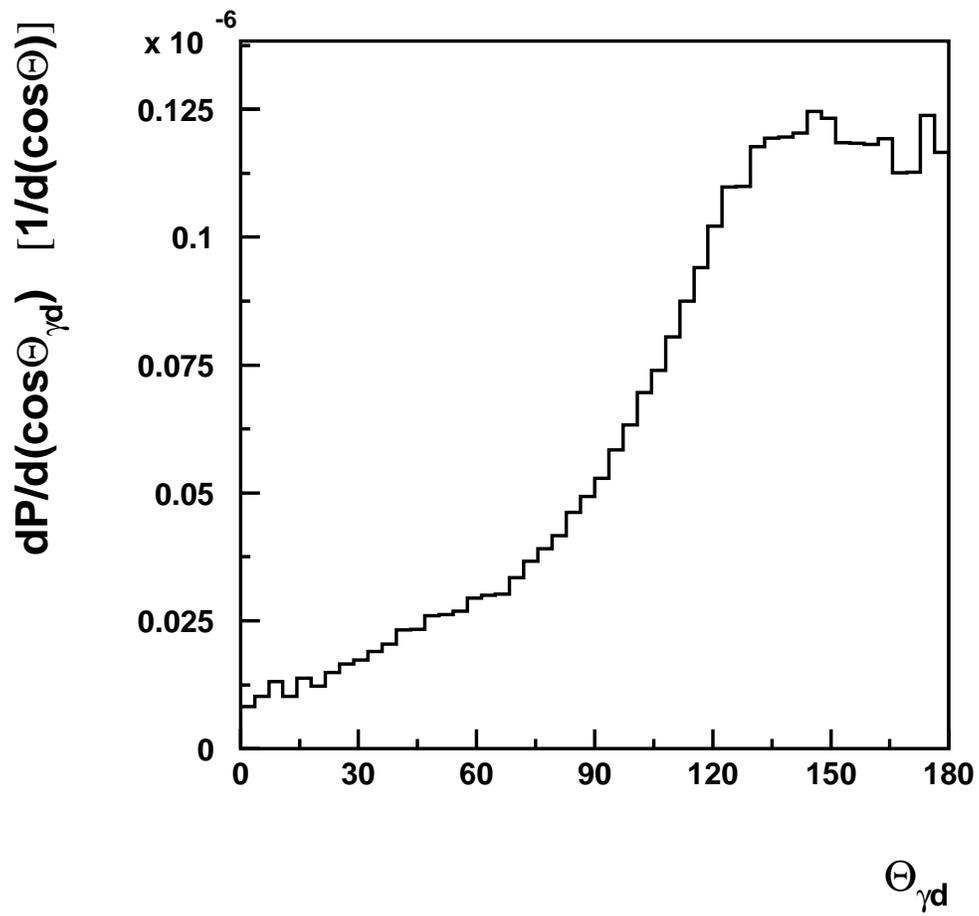,width=0.86\textwidth}
\vspace{0.5cm}
\caption{
Production probability
of a deuteron-photon pair
per participant collision as a function of the relative 
angle
$\Theta_{\gamma d}$~ at the beam energy of $90$MeV/A
($E_\gamma >30$MeV).}
\end{center}
\end{figure}

\newpage
\begin{figure}
\begin{center}
\epsfig{file=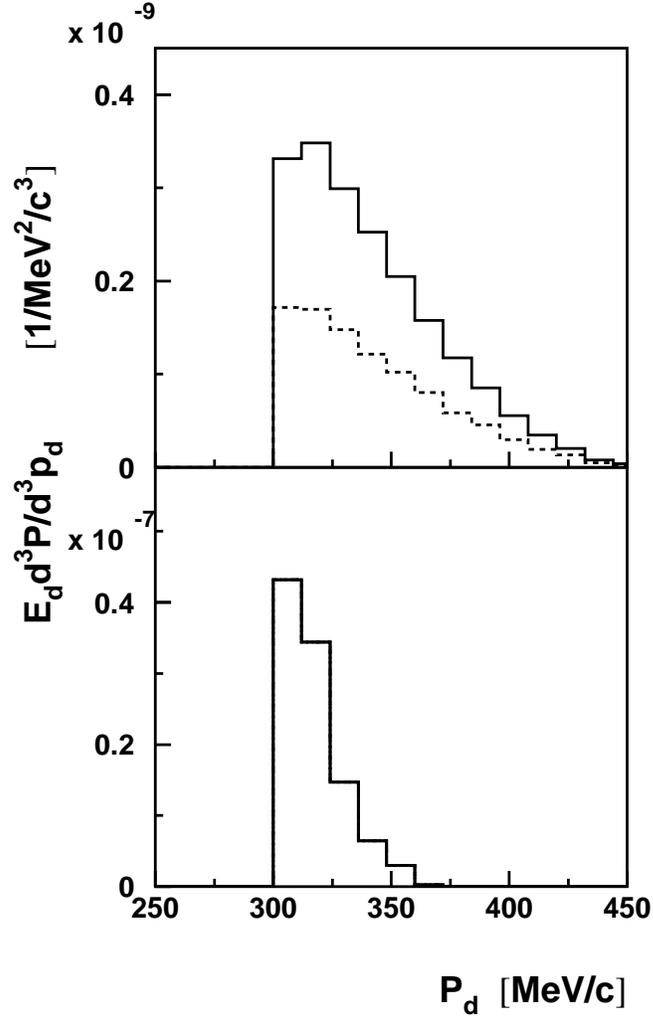,width=0.7\textwidth}
\vspace{0.5cm}
\caption{
Top shows
the probability density for producing a deuteron in
the 2-body process at $90^\circ$ in the nucleon-nucleon
c.m., as a function of the
deuteron momentum.  The solid line is obtained for the
same conditions as for Fig. 3.  The dashed line is
obtained with the supplementary condition that the 
energy of the
photon, in the reconstructed c.m.\ of the 
nucleon-nucleon
collision, is
larger than $60$MeV.  Bottom of the figure shows the
probability density for deuteron production
in the 3-body process
at $90^\circ$ in the nucleon-nucleon
c.m., shown
as a function of the deuteron momentum at the energy
$90MeV$/A.}
\end{center}
\end{figure}

\newpage
\begin{figure}
\begin{center}
\epsfig{file=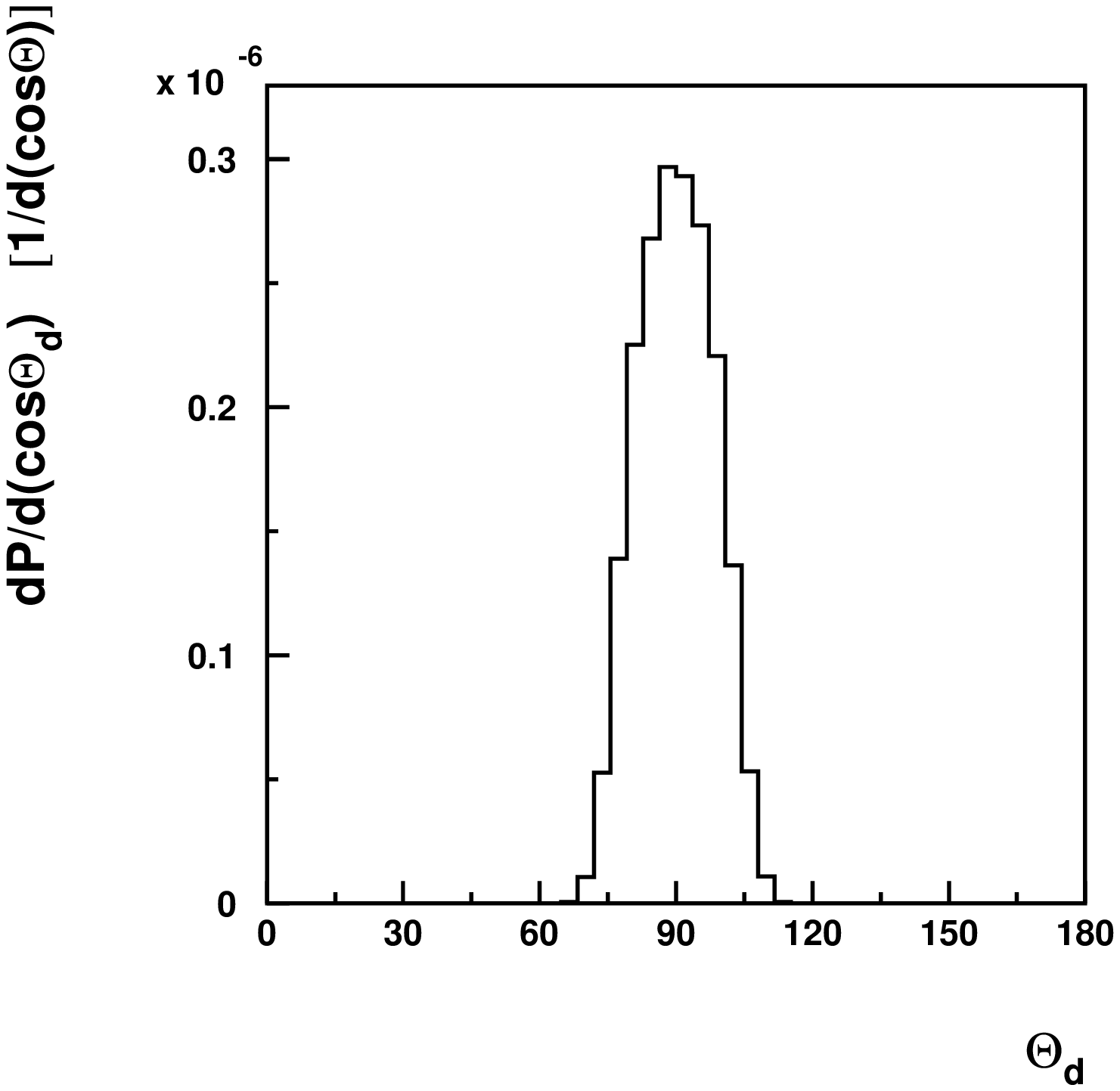,width=0.86\textwidth}
\vspace{0.5cm}
\caption{
Probability density for emitted deuterons as a function 
of
the angle in nucleus-nucleus c.m.\, for the same 
conditions as
in Fig. 3.
}
\end{center}
\end{figure}


\begin{thebibliography}{99}

\bibitem{gp} K. Gudima et al. (TAPS Collaboration), 
Phys. Rev. Lett.
{\bf 76} (1996) 2412.


\bibitem{ogmott} H.B Ghassib, R.F. Bishop and M.R. 
Strayer, J. Low
Temp. Phys. {\bf 20} (1976) 393; \\
G. R\"opke, L. M\"unchow and
H. Schulz, Nucl. Phys. {\bf A379} (1982) 536;    \\
B.E. Vonderfecht, C.C.
Gearht, W.H. Dickhoff, A. Polls and A. Ramos, Phys. 
Lett. {\bf B253}
(1991) 1.


\bibitem{ogmott1} M. Schmidt, G. R\"opke and H. Schulz, 
Annals
of Physics {\bf 202} (1990) 57.


\bibitem{bd} P. Danielewicz and G.F. Bertsch, Nucl. 
Phys {\bf A533}
(1991) 712.


\bibitem{review} H. Nifnecker and J.A. Pinston, Prog. 
Part. Nucl. Phys.
{\bf 23} (1989) 271;\\
W. Cassing, V. Metag, U. Mosel and K. Niita, Phys. Rep. 
{\bf 188}
(1990) 364.


\bibitem{bertsch77} G.F. Bertsch, Phys. Rev. {\bf C15} 
(1977) 713.


\bibitem{partovi} F. Partovi, Ann. Phys. (NY) {\bf 27} 
(1964) 79.


\bibitem{3n-dn} G. R\"opke and H. Schulz, Nucl. Phys. 
{\bf A477}
 (1988) 472.

\bibitem{gan} N. Gan, K.-T. Brinkmann, A.L. Caralay, 
B.J. Fineman,
W.J. Kernan, R.L. McGrath and P. Danielewicz,
Phys Rev. {\bf C49} (1994) 298.


\bibitem{rostdn} M. Beyer, G. R\"opke and A. Sedrakian, 
Phys. Lett.
{\bf B376} (1996) 7.

\end{thebibliography}
\end{document}